\documentclass[12pt,english,reqno]{smfart}
\usepackage{mathptmx}
\usepackage{epsfig}
\renewcommand{\theequation}{\arabic{section}.\arabic{equation}}
\theoremstyle{plain}
\newtheorem{thm}{Theorem}
\newtheorem{cor}{Corollary}
\theoremstyle{definition}
\newtheorem*{rem}{Remark}
\begin{document}
\title[Rigorous Hubbard-Stratonovich]
{Hyperbolic Hubbard-Stratonovich transformation made rigorous}
\author{Y.V.\ Fyodorov}
\address{School of Mathematical Sciences, University of Nottingham, UK}
\author{Y.\ Wei}
\address{School of Mathematical Sciences, University of Nottingham, UK}
\author{M.R.\ Zirnbauer}
\address{Institut f\"ur Theoretische Physik, Universit\"at zu K\"oln, Germany}
\date{January 11, 2008}
\begin{abstract}
We revisit a long standing issue in the theory of disordered electron
systems and their effective description by a non-linear sigma model:
the hyperbolic Hubbard-Stratonovich (HS) transformation in the
bosonic sector. For time-reversal invariant systems without spin this
sector is known to have a non-compact orthogonal symmetry $\mathrm{O}
_{p,\,q\,}$. There exists an old proposal by Pruisken and Sch\"afer
how to perform the HS transformation in an $\mathrm{O}_{p,\,q}
$-invariant way. Giving a precise formulation of this proposal we
show that the HS integral is a sign-alternating sum of integrals over
disjoint domains.
\end{abstract}
\maketitle

\section{Introduction}\label{sect:intro}
\setcounter{equation}{0}

Initiated by work of Wegner \cite{weg}, Sch\"afer \& Wegner
\cite{sw}, and Pruisken \& Sch\"{a}fer \cite{ps}, non-compact
non-linear sigma models as well as their supersymmetric
generalizations due to Efetov \cite{efe} have long been a standard
tool in the field of disordered electron physics. Areas of
application include single electron motion in disordered and chaotic
mesoscopic systems \cite{mirlin}, chaotic scattering \cite{vwz,fs},
localization and delocalization in systems exhibiting the Integer
Quantum Hall Effect \cite{QHE}, statistical properties of the Dirac
spectrum in non-abelian gauge field backgrounds \cite{vw}, to mention
only a few.

The utility of the non-linear sigma model derives from the fact that
it exposes the long-wavelength degrees of freedom of the disordered
system, which are hidden in the original microscopic formulation by a
Hamiltonian with quenched, i.e., time-indepen\-dent, random
parameters. These degrees of freedom correspond to interacting
diffusion modes, known to be responsible for the universal behavior
of spectral and eigenfunction statistics in a broad class of
disordered systems \cite{efe,mirlin}.

While the final results are expected to exhibit a high degree of
universality, the mathematical tools employed to derive the
non-linear sigma model may vary depending on the type of microscopic
model under consideration. A concise introduction to this issue can
be found in \cite{susymethod}; see also the recent papers
\cite{superbos1,superbos2}. For some types of microscopic model the
following mathematical identity plays a key role in the derivation:
\begin{equation}\label{eq:hs}
    C_0\; \mathrm{e}^{- \mathrm{Tr}\,{A}^2} = \int_D \mathrm{e}^{-
    \mathrm{Tr}\,{R}^2 - 2\mathrm{i}\, \mathrm{Tr}\, A R} |dR| \;,
\end{equation}
where $C_0$ is a constant and $|dR|$ denotes the Lebesgue measure of
a normed vector space of matrices $R\,$. In a situation with compact
orthogonal symmetry, which arises when the fermionic sector of the
theory is considered, one has license to take the integration domain
$D$ to be the real symmetric matrices $R\,$, and it is then trivial
to do the Gaussian integral by completing the square and shifting
variables. However, for reasons reviewed in Sect.\
\ref{sect:background} the bosonic sector (say, of time-reversal
invariant systems without spin) calls for $A$ to be a
\emph{non}-symmetric matrix composed of elements
\begin{equation}
    A_{ij} = \sum\nolimits_{a=1}^N \varphi_{i,\,a}
    \varphi_{j,\,a}\, s_j \;,
\end{equation}
with $(p+q)\times N$ real numbers $\varphi_{i,\,a}$ and $s_1 = \ldots
= s_p = - s_{p+1} = \ldots = - s_{p+q} = 1\,$.

The diagonal matrix $s$ with entries $s_{ij} = s_i\, \delta_{ij}$
determines a non-compact variant $\mathrm{O}_{p,\,q}$ of the real
orthogonal group by the condition $g^\mathrm{t} s g = s\,$, where
$g^\mathrm{t}$ means the transpose of the matrix $g$. First
discovered in the present context by Wegner \cite{weg}, the group
$\mathrm{O}_{p,\,q}$ is sometimes referred to as the `hyperbolic
symmetry' of the problem at hand.

In the situation with hyperbolic symmetry, choosing a good domain of
integration for $R$ and making rigorous sense of the integral
(\ref{eq:hs}) are non-trivial tasks. The difficulty arises from the
general context of the so-called Hubbard-Stratonovich method: the
exponential $\mathrm{e}^{- 2\mathrm{i} \,\mathrm{Tr}\, AR}$ must be
kept bounded -- or else the next step of the method, which is to
integrate over the microscopic fields $\varphi_{i,\,a}\,$, would be
invalid. Since the real matrix $A$ satisfies the symmetry relation $A
= s A^\mathrm{t} s$ one might think that one should try to integrate
over the domain of all real matrices $R$ subject to the symmetry $R =
s R^\mathrm{t} s$. Unfortunately, the resulting quadratic form
$\mathrm{Tr}\,R^2 = \mathrm{Tr} \, R\, s R^\mathrm{t} s$ is of
indefinite sign and therefore such a choice of integration domain for
$R$ makes the $R$-integral divergent.

Nonetheless, a valid solution of the problem posed by (\ref{eq:hs}),
i.e., a choice of integration domain for $R$ making the $R$-integral
converge while keeping the integrand bounded as a function of $A$,
was offered in the paper by Sch\"afer and Wegner \cite{sw}; by
deforming some of the real freedoms in $R$ into the complex numbers
they parameterize $R$ as
\begin{displaymath}
    R = P - \mathrm{i} \lambda T s T^{-1} \;,
\end{displaymath}
where $P$ runs through the real symmetric matrices that commute with
$s\,$, the constant $\lambda$ is any positive real number, and $T \in
\mathrm{SO}_{p,\,q} \,$. The reader is referred to the review
\cite{susymethod} for a detailed discussion. Now the Sch\"afer-Wegner
choice of domain has, besides its many merits, a certain drawback: it
lacks invariance under the action of the symmetry group $\mathrm{O}
_{p,\,q}\,$. This may be the reason why the Sch\"afer-Wegner choice
was never much in use by the disordered electron physics community.

\subsection{Statement of result}\label{sect:result}

Another choice of integration domain for (\ref{eq:hs}), which meets
the afore-mentioned requirements and also has the desirable property
of $\mathrm{O}_{p,\,q}$-invariance, first appeared in a paper by
Pruisken and Sch\"afer \cite{ps}. Beginning with \cite{vwz} this
choice came to be used in numerous later works. In the present paper
we are going to establish this so-called Pruisken-Sch\"afer domain as
a rigorous alternative to that of Sch\"afer-Wegner. There exists,
however, a mathematical subtlety: although our choice of domain
agrees with Pruisken-Sch\"afer as a union $D = \bigcup_\sigma
D_\sigma$ of \emph{sets} $D_\sigma\,$, it differs as an element $D =
\sum_\sigma \mathrm{sgn}(\sigma) D_\sigma$ of the vector space of
integration \emph{chains} (for the details, see below). Some
low-dimensional special cases of it were worked out in \cite{yf1,wy}.

The Pruisken-Sch\"afer domain, $D$, is an open subspace of the
$\frac{1}{2}(p+q)(p+q+1) $-dimensional space of real matrices $R$
obeying the mixed symmetry relation $R = s R^\mathrm{t} s$. Because
such matrices may in general have \emph{complex} eigenvalues and
eigenvectors, one defines $D$ as the subspace of matrices $R = s
R^\mathrm{t} s$ that can be diagonalized by conjugation $R \mapsto
g^{-1} R\, g$ with an element $g \in \mathrm{O}_{p,\,q}$ of the
\emph{real} (non-compact) orthogonal group. This set $D$ turns out to
be the union $D = \bigcup_\sigma D_\sigma$ of $\binom{p+q}{p} =
\binom{p+q}{q}$ sub-domains $D_\sigma \,$.

The domains $D_\sigma$ are enumerated as follows. With the exception
of cases forming a set of measure zero, an $\mathrm{O}_{p,\,q}
$-diagonalizable matrix $R = s R^\mathrm{t} s$ has $p$ `space-like'
and $q$ `time-like' eigenvalues $\lambda \in \mathbb{R}$ -- here we
use the language of relativity theory to communicate that the
corresponding eigenvector $v$ has the property $v^\mathrm{t} s v > 0$
or $v^\mathrm{t} s v < 0$, respectively. In the generic situation
without degeneracies, the eigenvalues of $R$ can be arranged in
decreasing order and this ordered sequence translates into a binary
sequence $\sigma$ by writing, say, the symbols $\bullet$ for
space-like and $\circ$ for time-like eigenvalues.

The binary sequence $\sigma$ encoding the relative order of
space-like and time-like eigenvalues is an $\mathrm{O}_{p,\,q}
$-invariant. Moreover, collisions between eigenvalues of space-like
and time-like type generically lead to the birth of complex
eigenvalues, thereby taking us outside the integration domain for
$R\,$. This means that the $\mathrm{O}_{p,\,q}$-diagonalizable
matrices $R$ of binary sequence $\sigma$ form an $\mathrm{O}_{p,\,q}
$-invariant domain, which we denote by $D_\sigma\,$, and any two such
domains $D_\sigma$ and $D_{\sigma^\prime}$ with $\sigma \not=
\sigma^\prime$ can only touch each other in subspaces of lower
dimension (more precisely, of codimension two).

The integration measure $|dR|$ is defined to be the flat one for all
domains $D_\sigma\,$:
\begin{equation}\label{eq:flat}
    |dR| = \prod\nolimits_{i \le j} \; dR_{ij} \;.
\end{equation}
Now the startling feature of the following statement is that the
integral on the right-hand side of (\ref{eq:hs}) is proposed to be a
sum of integrals $\int_{D_\sigma}$ with \emph{alternating} sign!
\begin{thm}\label{thm:1}
There exists some choice of cutoff function $R \mapsto
\chi_\varepsilon(R)$ (converging pointwise to unity as $\varepsilon
\to 0$), and a unique choice of sign function $\sigma \mapsto
\mathrm{sgn} (\sigma) \in \{ \pm 1 \}$ and a constant $C_{p,\,q}$
such that
\begin{equation}\label{eq:thm1}
    C_{p,\,q} \lim_{\varepsilon \to 0} \sum_\sigma \mathrm{sgn}
    (\sigma) \int_{D_\sigma} \mathrm{e}^{- \mathrm{Tr}\, R^2 -
    2\mathrm{i}\, \mathrm{Tr} \, AR} \chi_\varepsilon(R) |dR|
    = \mathrm{e}^{ - \mathrm{Tr}\, A^2}
\end{equation}
holds true for all matrices $A = s A^\mathrm{t} s$ with the
positivity property $A s > 0\,$.
\end{thm}
\begin{rem}
It will be shown that $\mathrm{sgn}(\sigma)$ is the parity of the
number of transpositions $\bullet \leftrightarrow \circ$ needed to
reduce the binary sequence $\sigma$ to the extremal form $\sigma_0 =
\bullet \cdots \bullet \circ \cdots \circ$. It should be emphasized
that (\ref{eq:thm1}) is \emph{not} a standard Gaussian integral; in
fact, we have not succeeded in finding a proof of this formula by
completing the square and shifting.
\end{rem}
It is informative to give an alternative expression for the integral
(\ref{eq:thm1}) in terms of the eigenvalues of $R\,$. Let $R = g
\lambda g^{-1}$ with $\lambda = \mathrm{diag}(\lambda_1 , \ldots,
\lambda_{p+q}) \in \mathbb{R}^{p+q}$ and $g \in \mathrm{SO}_{p,\,q}
\,$. The volume element transforms as $|dR| = J(\lambda) |d\lambda|
\, dg$ where $|d\lambda| = \prod_{i=1}^{p+q} d\lambda_i$ and $dg$ is
a positive Haar measure for the connected group $\mathrm{SO}_{p,\,q}
\,$. The Jacobian is
\begin{displaymath}
    J(\lambda) = \prod\nolimits_{i<j}\vert\lambda_i -\lambda_j\vert\;.
\end{displaymath}
\begin{cor}\label{cor:1}
Let a function $J^\prime(\lambda)$ with alternating sign on
$\mathbb{R}^{p+q}$ be defined by
\begin{displaymath}
    J^\prime(\lambda) = J(\lambda) \prod_{i=1}^p
    \prod_{j=p+1}^{p+q} \mathrm{sign}(\lambda_i-\lambda_j)\;.
\end{displaymath}
Then, assuming the positivity $A s > 0$ of the matrix $A = s
A^\mathrm{t} s\,$, we have
\begin{displaymath}
    C_{p,\,q} \lim_{\varepsilon \to 0} \int_{\mathbb{R}^{p+q}}
    \left( \int_{\mathrm{SO}_{p,\,q}} \mathrm{e}^{- 2\mathrm{i}
    \,\mathrm{Tr}\, A\, g \lambda g^{-1}} \chi_\varepsilon(g
    \lambda g^{-1})\, dg \right) \mathrm{e}^{- \mathrm{Tr}\,
    \lambda^2} J^\prime(\lambda) | d\lambda | = \mathrm{e}^{-
    \mathrm{Tr}\, A^2} \;.
\end{displaymath}
\end{cor}
\begin{rem}
The first $p$ eigenvalues $\lambda_j$ are space-like, the last $q$
time-like. To deduce Cor.\ \ref{cor:1} from Thm.\ \ref{thm:1}, we
need only observe that the function $J^\prime(\lambda)$ differs from
the Jacobian $J(\lambda)$ precisely by our sign function $\lambda
\mapsto \mathrm{sgn}(\sigma(\lambda))$:
\begin{displaymath}
    J^\prime(\lambda) = J(\lambda) \, \mathrm{sgn}(\sigma(\lambda))\;.
\end{displaymath}
The presence of this alternating sign factor was conjectured in
\cite{wy}; see also \cite{yf1}.

In previous work, beginning with \cite{vwz}, the identity of Cor.\
\ref{cor:1} was assumed to hold without the alternating sign. It
became apparent in 1995 that this assumption was unfounded
\cite{gossiaux} though it remained unclear at that time how to
correct the mistake. Problems with the Pruisken-Sch\"afer domain were
mentioned in \cite{zirn-sss}, but for various reasons the issue was
never much emphasized in the published literature. (For one thing,
all results derived using $J(\lambda)$ instead of $J^\prime(\lambda)$
stand correct after saddle-point approximation for large $N$, as the
integral in the large-$N$ limit is dominated by contributions from a
single domain $D_\sigma\,$. For another, the Sch\"afer-Wegner domain
was available \cite{zirn-sss} as a rigorous alternative in order to
do all intermediate steps correctly.) The present paper solves this
long standing problem in a satisfactory if perhaps surprising manner.

Let it be pointed out that the matrix $A s$ appearing in applications
is not strictly positive but positive {\it semi}-definite. This is,
however, a minor issue as an easy variant of the formula
(\ref{eq:thm1}) takes care of the semi-definite case; see \cite{yf1}
for the details.

For the related case of non-compact unitary symmetry $\mathrm{U}_{p,
\,q}$ a formula analogous to that of Cor.~\ref{cor:1} had been
established in \cite{yf1}, using the special feature of semiclassical
exactness by the Duistermaat-Heckman theorem \cite{fes}. Although it
will not be shown here, the methods of the present paper are robust
and can be adapted to handle the case of $\mathrm{U}_{p,\,q}$ as
well, without taking recourse to semiclassical exactness.
\end{rem}
The paper is organized as follows. In Sect.\ \ref{sect:background} we
present some background material concerning the Hubbard-Stratonovich
method. Basic results needed from integral calculus are collected in
Sect.\ \ref{sect:basics}. Then, in Sect.\ \ref{sect:1,1}, we work out
the simple but instructive case of $p = q = 1$ in detail. The case of
general $p$ and $q$ is handled in Sect.\ \ref{sect:p,q} by reduction
to $p = q = 1$.

\section{Background}\label{sect:background}
\setcounter{equation}{0}

To make the present paper self-contained, we now describe the steps
prior to (\ref{eq:thm1}), called the Hubbard-Stratonovich
transformation in the present context. Although the true power of
this transformation is to allow the treatment of non-trivial models
with \emph{local} $\mathrm{O}_N$ gauge symmetry \cite{sw} or even
without such symmetry \cite{ps}, we will restrict ourselves here to
reviewing the basic steps at the example of the simplest random
matrix model of this universality class: the Gaussian Orthogonal
Ensemble (GOE).

In the present work we are concerned with time-reversal invariant
systems without spin. For such systems, the quantum mechanics of
stationary states can be done over the field $\mathbb{R}$ of real
numbers. Assuming the real Hilbert space to be of finite dimension,
$N$, we may express the Hamiltonian operators $H$ by real symmetric
$N \times N$ matrices. The Gaussian Orthogonal Ensemble by definition
is an $\mathrm{O}_N$-invariant probability measure $d\mu(H)$ on the
space of such matrices. For our purposes, the GOE measure $d\mu(H)$
is best characterized by its Fourier transform:
\begin{equation}\label{eq:GOE}
    \left\langle \mathrm{e}^{\mathrm{i}\,
    \mathrm{Tr}\, HK} \right\rangle_\mathrm{GOE} \equiv \int
    \mathrm{e}^{\mathrm{i}\, \mathrm{Tr}\, HK} d\mu(H) =
    \mathrm{e}^{-\frac{b^2}{2N}\mathrm{Tr}\, K^2} \;,
\end{equation}
where the Fourier variable $K$ is any symmetric matrix, and $b \in
\mathbb{R}$ is a parameter.

An important object of the theory (see, e.g., \cite{susymethod}) is
the expectation value of the reciprocal of a product of (square roots
of) characteristic polynomials of $H:$
\begin{equation}\label{2}
    F(z_1, \ldots, z_{p+q}) := \left\langle\prod\nolimits_{j = 1}
    ^{p+q}\; \mathrm{Det}^{-1/2} (z_j-H) \right\rangle_\mathrm{GOE}
    \quad (z_j \in \mathbb{C}\setminus\mathbb{R}) \;.
\end{equation}
We assume that $\mathfrak{Im}\, z_j > 0$ for $j = 1, \ldots, p$ and
$\mathfrak{Im}\, z_j < 0 $ for $j = p+1 , \ldots, p+q\,$. To compute
such an expectation value, we use the trick of representing the
reciprocal square root of each determinant as a Gaussian integral
over vector variables $\varphi:$
\begin{displaymath}
    \mathrm{Det}^{-1/2} (z_j-H) = (\mathrm{i} \pi s_j)^{-N/2}
    \int_{\mathbb{R}^N} \mathrm{e}^{\mathrm{i} s_j\, ( \varphi ,
    \; \varphi \, z_j - H \varphi)} |d\varphi| \;,
\end{displaymath}
where $|d\varphi| = \prod_{a=1}^N d\varphi_a$ and the factor $s_j :=
\mathrm{sign}(\mathfrak{Im}\, z_j)$ ensures the convergence of the
integral. $(\phi,\psi) := \sum_{a=1}^N \phi_a \psi_a$ means the
Euclidean scalar product of $\mathbb{R}^N$.

Substituting such integrals into (\ref{2}), one immediately performs
the GOE average by employing the characteristic property
(\ref{eq:GOE}). The result can then be expressed in terms of a matrix
$A( \varphi)$ of size $(p+q)\times (p+q)$ with entries $A(\varphi
)_{ij} = \sum_{a=1}^N \varphi_{i,\,a} \varphi_{j,\,a\,} s_j :$
\begin{equation}\label{4}
    F(z_1,\ldots,z_{p+q}) = \int \mathrm{e}^{\mathrm{i}
    \sum_{j=1}^{p+q} s_j \, z_j\, (\varphi_j\,,\,\varphi_j)}
    \mathrm{e}^{-\frac{b^2}{2N}\mathrm{Tr}\, A(\varphi)^2}
    \prod_j \frac{|d\varphi_j|}{(\mathrm{i}\pi s_j)^{N/2}} \;.
\end{equation}

Here is where the identity (\ref{eq:hs}) comes in: it enables us to
linearize the quadratic term $\mathrm{Tr}\, A(\varphi)^2$ in the
exponent, thereby reducing the integral over the $N$-component
vectors $\varphi_1 , \ldots, \varphi_{p+q}$ to $N$ decoupled Gaussian
integrals, one for each component $a = 1, \ldots, N$. When the latter
integrals are carried out, (\ref{4}) becomes an integral over the
collective variables $R$ only. Using the rigorous form
(\ref{eq:thm1}) of the relation (\ref{eq:hs}) we obtain
\begin{equation}\label{eq:intrep}
    F(z) = C_{p,\,q}^\prime \sum_\sigma \mathrm{sgn}(\sigma)
    \int_{D_\sigma} \mathrm{e}^{-\frac{N}{2b^2} \mathrm{Tr}\,R^2}
    \mathrm{Det}^{-N/2} (z - R)\, | dR | \;,
\end{equation}
where $z := \mathrm{diag}(z_1,\ldots,z_{p+q})$ and, with the
assumption that $N$ is large enough, the cutoff function
$\chi_\varepsilon$ has been removed by sending the regularization
parameter $\varepsilon$ to zero.

The integral representation (\ref{eq:intrep}) for $F(z)$ is well
suited for saddle-point analysis in the limit of large $N$. We can
now appreciate why the identity (\ref{eq:hs}) is useful here: it
serves the purpose of exposing the good variables in which to perform
the large-$N$ limit. Let this fact suffice as a motivation for our
labors in the body of this paper.

\section{Basics from calculus}\label{sect:basics}
\setcounter{equation}{0}

Recalling from Sect.\ \ref{sect:result} the definition of the
connected domains $D_\sigma$ (cf.\ also below) we consider the
following alternating sum of integrals:
\begin{equation}\label{eq:sum1}
    I(A) = \lim_{\varepsilon\to 0+} \sum_\sigma \mathrm{sgn}(\sigma)
    \int_{D_\sigma} \mathrm{e}^{- \mathrm{Tr}\, (R + \mathrm{i}A)^2}
    \chi_\varepsilon (R + \mathrm{i}A)\, |dR| \;,
\end{equation}
where $\chi_\varepsilon$ is some smooth cutoff function which
regularizes the integral and converges pointwise to unity in the
limit $\varepsilon \to 0\,$. The real matrix $A$ is subject to the
conditions $A = s A^\mathrm{t} s$ and $A s > 0\,$. Choosing the
Lebesgue measure $|dR|$ as in (\ref{eq:flat}) we shall prove that the
integral $I(A)$ does not depend on the matrix $A$, or equivalently,
that all of its directional derivatives vanish:
\begin{equation}\label{eq:indep}
    \frac{d}{dt} I(A + t \dot{A}) \Big\vert_{t = 0} = 0 \;.
\end{equation}
Once (\ref{eq:indep}) has been established, the Hubbard-Stratonovich
transformation (\ref{eq:thm1}) follows by multiplying both sides of
(\ref{eq:sum1}) with a constant $\mathrm{e}^{- \mathrm{Tr}\, A^2}$.

To prepare our treatment, we recall a few basic facts from calculus.
Given a vector field $v$ on an $n$-dimensional differentiable
manifold $M$, let $[-\delta , \delta] \ni t \mapsto \phi_t$ be the
flow of $v$, i.e., the one-parameter family of mappings $\phi_t : \,
M \to M$ determined by $\phi_{t=0}(p) = p$ and $\frac{d}{dt}
\phi_t(p) = v(\phi_t(p))$ for all $p \in M$. In local coordinates
$x^1, \ldots, x^n$, we have
\begin{displaymath}
    \frac{d}{dt} x^i \circ \phi_t = v^i \circ \phi_t
    \qquad (v = v^i \, \partial / \partial x^i) \;.
\end{displaymath}
We are using the summation convention: an index appearing twice (once
as a covariant and once as a contravariant index) is understood to be
summed over.

For a differential form $\alpha = \alpha_{i_1 \ldots \, i_k} \,
dx^{i_1} \wedge \cdots \wedge dx^{i_k}$ on $M$ let
\begin{displaymath}
    \phi_t^\ast \alpha = (\alpha_{i_1 \ldots \, i_k} \circ
    \phi_t)\; d(x^{i_1} \circ \phi_t) \wedge \cdots \wedge d(x^{i_k}
    \circ \phi_t)
\end{displaymath}
denote the pullback of $\alpha$ by $\phi_t\,$. The Lie derivative of
$\alpha$ w.r.t.\ the vector field $v$ is then defined by
differentiation at $t = 0 :$
\begin{displaymath}
    \mathcal{L}_v \alpha = \frac{d}{dt} \phi_t^\ast \alpha
    \Big\vert_{t = 0} \;.
\end{displaymath}
By the relation $\phi_t^\ast(\alpha \wedge \beta) = (\phi_t^\ast
\alpha) \wedge (\phi_t^\ast \beta)$, the Lie derivative satisfies the
Leibniz rule $\mathcal{L}_v (\alpha \wedge \beta) = (\mathcal{L}_v
\alpha) \wedge \beta + \alpha \wedge (\mathcal{L}_v \beta)$. Cartan's
formula expresses it as
\begin{displaymath}
    \mathcal{L}_v = d \circ \iota_v + \iota_v \circ d \;,
\end{displaymath}
where $d$ is the exterior derivative,
\begin{displaymath}
    d \left( \alpha_{i_1 \ldots \, i_k}\, dx^{i_1} \wedge \cdots
    \wedge dx^{i_k} \right) = d\alpha_{i_1 \ldots \, i_k} \wedge
    dx^{i_1} \wedge \cdots \wedge dx^{i_k} \;,
\end{displaymath}
and $\iota_v$ denotes the operation of contraction with the
vector field $v :$
\begin{displaymath}
    (\iota_v \alpha)_{i_1 \ldots \, i_{k-1}} = v^i \left( \alpha_{i\,
    i_1 \ldots \, i_{k-1}} - \alpha_{i_1 i\, \ldots \, i_{k-1}} +
    \ldots + (-1)^{k-1} \alpha_{i_1 \ldots \, i_{k-1} i} \right) \;.
\end{displaymath}

After this summary of basic facts, let $\omega$ be an $n$-form on $M$
and consider the integral $\int_M (\mathcal{L}_v f) \omega$ where
$\mathcal{L}_v f = \iota_v \, df$ is the derivative of some function
$f : \, M \to \mathbb{C}$ in the direction of the vector field $v$.
Assuming that $\omega$ is invariant by the flow of $v$, the Leibniz
rule gives
\begin{displaymath}
    \int_M (\mathcal{L}_v f) \omega = \int_M \mathcal{L}_v
    (f \omega) = \int_M (d \circ \iota_v) (f \omega) \;.
\end{displaymath}
By Stokes' theorem for the integral of the exact $n$-form $d(f \,
\iota_v \omega)$ it follows that
\begin{equation}\label{eq:IbP}
    \int_M (\mathcal{L}_v f) \omega = \int_{\partial M} f \, \iota_v
    \omega \;,
\end{equation}
where $\partial M$ is the $(n-1)$-dimensional boundary of $M$.

In what follows, we will employ (\ref{eq:IbP}) in the closely related
case where one is integrating against a density $| \omega |$ instead
of an $n$-form $\omega$. Formula (\ref{eq:IbP}) still holds true in
that case -- assuming of course that $\mathcal{L}_v |\omega| = 0\,$.
Note also that in order to integrate the twisted $(n-1)$-form $f\,
\iota_v |\omega|$, the boundary $\partial M$ must be equipped with an
\emph{outer} orientation, i.e., a transverse vector field pointing,
say, from the inside to the outside of $\partial M$. (Technically
speaking, a density $| \omega |$ on $M$ is a top-degree differential
form $\omega$ tensored by a section $s$ of the orientation line
bundle of $M$, and a choice of splitting $| \omega | = \omega \otimes
s$ determines an isomorphism between outer and inner orientations for
$\partial M$.)

To apply the formula (\ref{eq:IbP}) to the situation at hand, let
$\tau(\dot R) = \dot{R}_{ij}\, \partial / \partial R_{ij}$ denote the
vector field generating translations $\phi_t(R) = R + t \dot{R}\,$.
Then
\begin{displaymath}
    \frac{d}{dt} I(A+t\dot{A}) \Big\vert_{t=0} = \lim_{\varepsilon
    \to 0+} \mathrm{i} \sum_\sigma \mathrm{sgn}(\sigma) \int_{D_\sigma}
    \mathcal{L}_{\tau(\dot{A})} \left( \mathrm{e}^{- \mathrm{Tr}\, (R+
    \mathrm{i}A)^2} \chi_\varepsilon (R + \mathrm{i}A) \right) |dR| \;,
\end{displaymath}
where $\tau(\dot A) = \dot{A}_{ij}\,\partial / \partial R_{ij}\,$.
Here, recognizing the fact that our integrand depends on $R$ and $A$
only through the combination $R + \mathrm{i}A\,$, we have used the
identity
\begin{displaymath}
    \frac{\partial}{\partial A_{ij}} f(R + \mathrm{i} A) =
    \mathrm{i}\frac{\partial}{\partial R_{ij}} f(R+\mathrm{i}A)\;.
\end{displaymath}

Now the Lebesgue measure (or positive density) $|dR|$ is invariant
under translations $R \mapsto R + t \dot{A}\,$. Therefore
$\mathcal{L}_{\tau (\dot{A})} |dR| = 0\,$, and we may apply formula
(\ref{eq:IbP}) to obtain
\begin{equation}\label{eq:dIda}
    \frac{d}{dt} I(A + t \dot{A}) \Big\vert_{t = 0} = \lim_{
    \varepsilon \to 0+} \mathrm{i}\sum_\sigma \mathrm{sgn}(\sigma)
    \int_{\partial D_\sigma}\mathrm{e}^{- \mathrm{Tr}\, (R +
    \mathrm{i}A)^2} \chi_\varepsilon (R+\mathrm{i}A)
    \iota_{\tau(\dot{A})} |dR| \;.
\end{equation}

The main achievement of this paper is a proof that the alternating
sum of integrals on the right-hand side vanishes in the limit
$\varepsilon \to 0\,$. For that purpose we need to develop a good
understanding of the boundary components $\partial D_\sigma$ and the
twisted form $\iota_{\tau(\dot{A})} |dR|$ restricted to them. As a
first step, we illustrate the essential features of the argument at
the example of the symmetry group being $\mathrm{O}_{1 ,1}\,$. This
example plays a very important role as the general case will be
handled by reduction to it.

\section{The case of $\mathrm{O}_{1,1}$-symmetry}
\label{sect:1,1}\setcounter{equation}{0}

Here we are going to deal with the special case of real $2 \times 2$
matrices $R$ subject to the linear symmetry relation
\begin{displaymath}
    R = s R^\mathrm{t} s \;, \quad s = \mathrm{diag}\,(1 , -1) \;,
\end{displaymath}
where $R \mapsto R^\mathrm{t}$ means the operation of taking the
matrix transpose. Such matrices can be parameterized by three real
variables $r_{11}\,$, $r_{12}$ and $r_{22}$ as
\begin{displaymath}
    R = \begin{pmatrix} r_{11} &r_{12} \\ -r_{12} &r_{22}
    \end{pmatrix} \;.
\end{displaymath}
Thus our space of matrices $R$ is isomorphic as a vector space to
$\mathbb{R}^3$. The measure of integration is the positive density
$|dR| = dr_{11}\, dr_{12}\, dr_{22}\,$. The matrix $A$ is of the same
form as $R$ but its matrix elements are constrained by
\begin{displaymath}
    a_{11} > 0 > a_{22} \;, \quad |a_{12}| < \sqrt{- a_{11}
    a_{22}}\;, \quad A = \begin{pmatrix} a_{11} & a_{12} \\
    -a_{12} & a_{22} \end{pmatrix} \;.
\end{displaymath}

Another way of describing the setting is to equip the real vector
space $\mathbb{R}^2$ with the sign-indefinite bilinear form $B$
determined by $s:$
\begin{displaymath}
    B(u,v) \equiv u^\mathrm{t} s v := u_1 v_1 - u_2 v_2 \;.
\end{displaymath}
The matrices $R$ are then symmetric with respect to $B$ in the sense
that
\begin{displaymath}
    B(u,R\,v) = B(R\,u,v)
\end{displaymath}
for all $u,v \in \mathbb{R}^2$. We denote the linear space of real
matrices $R$ with this symmetry property by $\mathrm{Sym}_B(
\mathbb{R}^2)$. Since $B$ has signature $(1,1)$, the symmetry group
of $B$ is the non-compact real orthogonal group $\mathrm{O}_{1,1}\,$.
Elements $g \in \mathrm{O}_{1,1}$ satisfy the equation
\begin{displaymath}
    s = g^\mathrm{t} s g \;,
\end{displaymath}
which is equivalent to saying that $B(u,v) = B(gu,gv)$ for all
$u,v\in \mathbb{R}^2$.

For reasons indicated in the Introduction (see Sect.\
\ref{sect:intro}) -- let us recall that in order for the integral
(\ref{eq:hs}) to exist one needs the inequality $\mathrm{Tr}\, R^2 >
0\,$, at least asymptotically -- we want our matrices $R \in
\mathrm{Sym}_B( \mathbb{R}^2)$ to be diagonalizable by the action $R
\mapsto g R\, g^{-1}$ of the real group $\mathrm{O}_{1,1}$ by
conjugation. The condition of diagonalizability is formulated most
clearly by expressing $R$ in a basis of light-like vectors of
$\mathbb{R}^2$:
\begin{displaymath}
    e_+ := \begin{pmatrix} 1 \\ 1 \end{pmatrix} \;, \quad
    e_- := \begin{pmatrix} 1 \\ -1 \end{pmatrix} \;,
\end{displaymath}
which obey $B(e_+ \,, e_+) = B(e_- \,, e_-) = 0$ and $B(e_+ \,, e_-)
= 2\,$. Applying $R$ to this basis one has
\begin{displaymath}
    R\, e_+ = \lambda e_+ + \eta e_- \;, \qquad
    R\, e_- = \lambda e_- + \xi e_+ \;,
\end{displaymath}
where $\lambda = \frac{1}{2} (r_{11} + r_{22})$, $\eta = \frac{1}{2}
(r_{11} - r_{22}) + r_{12}\,$, and $\xi = \frac{1}{2}(r_{11} -
r_{22}) - r_{12}\,$. Since the diagonal piece of $R$ in this basis is
a multiple of unity, the condition for diagonalizability is that the
product of off-diagonal matrix elements be positive: $\xi\eta > 0\,$.

The domain $\xi \eta > 0$ of $\mathrm{O}_{1,1}$-diagonalizability of
our matrices $R \in \mathrm{Sym}_B(\mathbb{R}^2)$ consists of two
connected components:
\begin{displaymath}
    D_{\bullet\circ} \,: \,\, \xi , \eta > 0 \;,
    \qquad D_{\circ\bullet}\,: \,\, \xi , \eta < 0 \;.
\end{displaymath}
The motivation for the notation ($\bullet, \circ$) will become clear
shortly. Here we simply remark that if $v_\bullet, v_\circ$ denotes a
pair of $R$-eigenvectors with $B(v_\bullet,v_\bullet) > 0 > B(v_\circ
, v_\circ)$ and $\lambda_\bullet, \lambda_\circ$ are the associated
eigenvalues, then $\lambda_\bullet > \lambda_\circ$ for $R \in
D_{\bullet\circ}$ and $\lambda_\circ > \lambda_\bullet$ for $R \in
D_{\circ\bullet}\,$.

Next, let $G_{A,\varepsilon}$ denote the regularized Gaussian
integrand
\begin{displaymath}
    G_{A,\varepsilon}(R):= \mathrm{e}^{-\mathrm{Tr}\,(R+\mathrm{i}
    A)^2} \chi_\varepsilon (R + \mathrm{i}A) \;,
\end{displaymath}
where we choose the cutoff function to be
\begin{displaymath}
    \chi_\varepsilon(R) := \mathrm{e}^{- \frac{\varepsilon}{2}
    \mathrm{Tr}\,(sR-Rs)^2}= \mathrm{e}^{-4\varepsilon\,{r_{12}}^2}\;.
\end{displaymath}
Using the coordinates $\lambda, \xi, \eta$ we have the expression
$G_{A,\varepsilon} = \mathrm{e}^{- f_2 - f_1 - f_0}$ with
\begin{eqnarray*}
    &&f_0 = - (a_{11}^2 + a_{22}^2) + 2 a_{12}^2 (1 - 2 \varepsilon)\;,
    \quad f_2 = 2 (\lambda^2+\xi\eta) + \varepsilon (\eta-\xi)^2\;,\\
    &&f_1 = 2\mathrm{i}(a_{11}+a_{22})\lambda+\mathrm{i}(a_{11}-a_{22})
    (\xi + \eta) + 2\mathrm{i} a_{12} (1-2\varepsilon)(\xi - \eta)\;.
\end{eqnarray*}
Let us observe that the integral $I(A)$ defined in (\ref{eq:sum1})
can now be written as
\begin{displaymath}
    I(A) = \lim_{\varepsilon \to 0+} \left( \int_{D_{\bullet\circ}}
    - \int_{D_{\circ\bullet}} \right) G_{A,\varepsilon}\, |dR| \;.
\end{displaymath}

\subsection{Description of boundaries}\label{sect:4.1}

\begin{figure}
    \begin{center}
        \epsfig{file=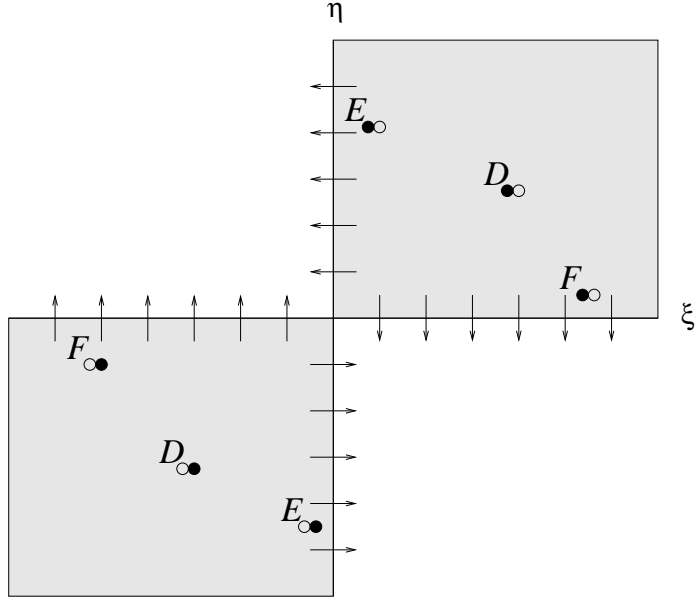,height=8cm}
    \end{center}
    \caption{Each sub-domain $D_\sigma$ has two boundary components,
    $E_\sigma$ and $F_\sigma\,$. The outer orientation
    of these boundaries is determined by the rule of crossing the
    boundary ``from the inside to the outside''.} \label{fig:2dom}
\end{figure}

By their definition via inequalities, the domains $D_\sigma$ (for
$\sigma = \bullet\circ , \circ\bullet$) are open in the
three-dimensional space $\mathrm{Sym}_B( \mathbb{R}^2)$ and have
boundaries $\partial D_\sigma$ which are two-dimensional. Each of the
boundaries $\partial D_\sigma$ is a union of 2 half-planes. To
describe this in detail, let 4 half-planes $E_\sigma$ and $F_\sigma$
be defined by
\begin{eqnarray*}
    &&E_{\bullet\circ} \, : \,\, \xi = 0 \;, \; \eta > 0 \;,
    \qquad F_{\bullet\circ} \, : \,\, \eta = 0 \;, \; \xi > 0\;,
    \\ &&E_{\circ\bullet} \, : \,\, \xi = 0 \;, \; \eta < 0\;,
    \qquad F_{\circ\bullet} \, : \,\, \eta = 0 \;, \; \xi < 0 \;.
\end{eqnarray*}
$E_\sigma$ and $F_\sigma$ make up the boundary of $D_\sigma$ (for
both $\sigma = \bullet\circ$ and $\sigma = \circ\bullet$). Being part
of $\partial D_{\circ\bullet}\,$, the half-plane $E_{\circ\bullet}$
inherits from $D_{\circ\bullet}$ an outer orientation by the
transverse vector field $\partial_\xi \equiv \partial / \partial \xi
\,$. (Indeed, starting from any point $R \in D_{\circ\bullet}$ very
close to the boundary $E_{\circ\bullet}$ and making a small step in
the direction of $\partial_\xi\,$, one crosses $E_{\circ\bullet}\,$.)
In the same sense, $F_{\circ\bullet} \subset \partial D_{\circ
\bullet}$ is oriented by $\partial_\eta$ while $E_\sigma \subset
\partial D_\sigma$ and $F_\sigma \subset \partial D_\sigma$ for
$\sigma = \bullet\circ$ are oriented by the opposite vector fields
$-\partial_\xi$ and $- \partial_\eta$ respectively; see Fig.\
\ref{fig:2dom}.

With these orientations being understood, we have the relations
\begin{displaymath}
    \partial D_\sigma = E_\sigma + F_\sigma \quad (\sigma =
    \bullet\circ \;, \circ\bullet) \;,
\end{displaymath}
which are equalities in the sense of chains, or distributions. (To
achieve equality in the sense of sets, we would have to include the
real line $\xi = \eta = 0\,$, which is part of both $\partial
D_{\bullet\circ}$ and $\partial D_{\circ\bullet}\,$. However, our aim
is to integrate bounded differential forms, and for that purpose the
lower-dimensional parts of the boundary are of no concern.)

With $\partial D_\sigma$ as described above, formula (\ref{eq:dIda})
now reads
\begin{equation}\label{eq:2.4}
    \frac{d}{dt} I(A+t\dot{A}) \Big\vert_{t=0} = \lim_{\varepsilon\to 0+}
    \left(\int_{\partial D_{\bullet\circ}} - \int_{\partial D_{\circ\bullet}}
    \right) \mathrm{i} G_{A,\varepsilon}\, \iota_{\tau(\dot{A})} |dR| \;.
\end{equation}

\subsection{Reorganization of boundary pieces}\label{sect:4.2}

\begin{figure}
    \begin{center}
        \epsfig{file=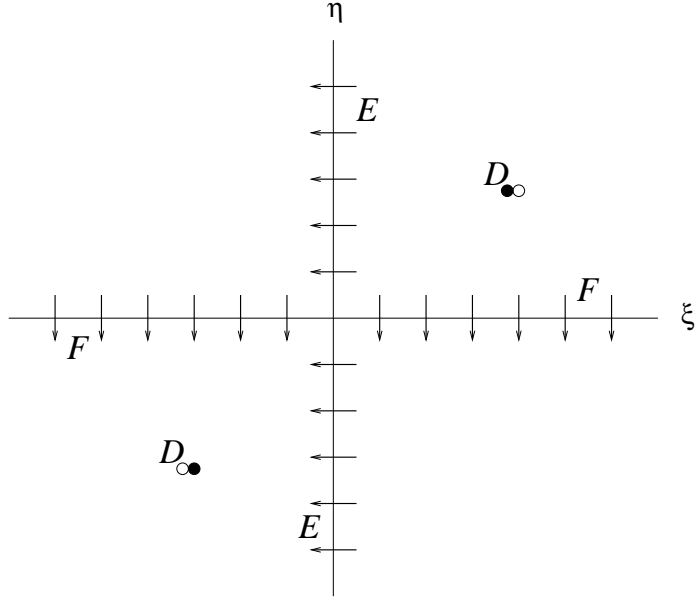,height=8cm}
    \end{center}
    \caption{Reorganization of the boundary pieces: defining
    $E = E_{\bullet\circ} - E_{\circ\bullet}$ and
    $F = F_{\bullet\circ} - F_{\circ\bullet}$ one obtains two
    consistently oriented 2-planes $E$ and $F$.}
    \label{fig:Or}
\end{figure}

The key idea is to reorganize the boundary half-planes into
consistently oriented closed manifolds. Define the full planes
\begin{eqnarray*}
    &&E \,:\,\, \xi = 0\quad\text{(oriented by }- \partial_\xi)\;,\\
    &&F \,:\,\,\eta = 0\quad\text{(oriented by }- \partial_\eta )\;,
\end{eqnarray*}
so that we have
\begin{displaymath}
    E = E_{\bullet\circ} - E_{\circ\bullet} \;, \quad
    F = F_{\bullet\circ} - F_{\circ\bullet} \;,
\end{displaymath}
in the sense of chains. Since
\begin{displaymath}
    \partial D_{\bullet\circ} - \partial D_{\circ\bullet} =
    E_{\bullet\circ} + F_{\bullet\circ} - E_{\circ\bullet}
    - F_{\circ\bullet} = E + F
\end{displaymath}
(still as chains), we obtain the following relation between
integrals:
\begin{displaymath}
    \int_{\partial D_{\bullet\circ}} \omega -
    \int_{\partial D_{\circ\bullet}} \omega =
    \int_E \omega + \int_F \omega
\end{displaymath}
for any twisted 2-form $\omega$ which is bounded (if not continuous
or smooth). In particular, this relation holds for $\omega =
\mathrm{i} G_{A,\varepsilon}\, \iota_{\tau(\dot{A})} |dR|$. Thus
\eqref{eq:2.4} turns into
\begin{equation}\label{eq:2.5}
    \frac{d}{dt} I(A + t\dot{A}) \Big\vert_{t = 0} =
    \lim_{\varepsilon\to 0+} \left( \int_E + \int_F \right)
    \mathrm{i} G_{A,\varepsilon}\,\iota_{\tau(\dot{A})}|dR|\;.
\end{equation}
A graphical sketch of the situation is shown in Fig.\ \ref{fig:Or}.

\subsection{Computation of boundary integral}\label{sect:4.3}

We are now going to show that the integral $\int_E \omega$ vanishes
for $\omega = \mathrm{i} G_{A,\varepsilon}\, \iota_{\tau( \dot{A})}
|dR|$ in the limit $\varepsilon \to 0\,$. Essentially the same
calculation shows that $\int_F \omega$ vanishes in this limit.

To handle the case of $E$, let $\psi :\, E \to \mathrm{Sym}_B(
\mathbb{R}^2)$ be the identical embedding and introduce coordinates
$\lambda, \eta$ on $E$ so that
\begin{displaymath}
    \psi^\ast \lambda = \lambda \;, \quad \psi^\ast \xi = 0 \;, \quad
    \psi^\ast \eta = \eta \;.
\end{displaymath}
Recalling $G_{A,\varepsilon} = \mathrm{e}^{-f_2 - f_1 - f_0}$ we pull
back the functions $f_2$ and $f_1$ to $E$:
\begin{displaymath}
    \psi^\ast f_2 = 2 \lambda^2 + \varepsilon \eta^2 \;, \quad
    \psi^\ast f_1 = 2\mathrm{i}(a_{11} + a_{22}) \lambda + \mathrm{i}
    (a_{11} - a_{22} - 2a_{12} (1-2\varepsilon)) \eta \;.
\end{displaymath}

Next we compute the pullback of the twisted 2-form $\iota_{\tau
(\dot{A})} |dR|\,$. For this, we write
\begin{displaymath}
    |dR| = dr_{11} \, dr_{12} \, dr_{22} = d\lambda \, d\xi \,
    d\eta := (d\lambda \wedge d\xi \wedge d\eta) \otimes \mathrm{Or}\;,
\end{displaymath}
where $\mathrm{Or} = + 1$ when $\mathrm{Sym}_B(\mathbb{R}^2)$ is
oriented by the ordered set of linear coordinates $\lambda$, $\xi$,
$\eta$ (or any even permutation thereof), and $\mathrm{Or} = -1$ when
the ordering $\lambda$, $\eta$, $\xi$ (or any other odd permutation
of $\lambda$, $\xi$, $\eta$) is chosen. Using the coordinate
expression of the vector field $\tau(\dot{A})$,
\begin{align*}
    \tau(\dot{A}) &= \dot{a}_{11} \frac{\partial}{\partial r_{11}} +
    \dot{a}_{12} \frac{\partial}{\partial r_{12}} + \dot{a}_{22}
    \frac{\partial}{\partial r_{22}}\;, \\ &=
    {\textstyle{\frac{1}{2}}} (\dot{a}_{11} + \dot{a}_{22})
    \partial_\lambda + {\textstyle{\frac{1}{2}}} (\dot{a}_{11} -
    \dot{a}_{22} - 2\dot{a}_{12}) \partial_\xi
    + {\textstyle{\frac{1}{2}}} (\dot{a}_{11} -
    \dot{a}_{22} + 2\dot{a}_{12}) \partial_\eta \;,
\end{align*}
we then obtain
\begin{displaymath}
    \psi^\ast \iota_{\tau(\dot{A})}|dR| ={\textstyle{\frac{1}{2}}}
    (\dot{a}_{11} - \dot{a}_{22} - 2 \dot{a}_{12}) \, (d\eta \wedge
    d\lambda) \otimes \mathrm{Or} \;.
\end{displaymath}
Now the presence of the orientation factor $\mathrm{Or}$ tells us
that an outer orientation by $\pm \partial_\xi$ translates into an
inner orientation by $\pm d\eta \wedge d\lambda$. Since our
$\lambda\eta$-plane $E$ is oriented by $- \partial_\xi\,$, we must
assign to it the orientation given by $- d\eta \wedge d\lambda\,$. In
this way, assembling everything we arrive at the expression
\begin{eqnarray*}
    &&\int_E \psi^\ast G_{A,\varepsilon}\, \iota_{\tau(\dot{A})}
    |dR| = - {\textstyle{\frac{1}{2}}} (\dot{a}_{11} - \dot{a}_{22}
    - 2 \dot{a}_{12}) \, \mathrm{e}^{a_{11}^2 + a_{22}^2 - 2 a_{12}^2
    (1-2\varepsilon)} \times \\ &&\times \int_\mathbb{R}
    \mathrm{e}^{-2 \lambda^2 - 2\mathrm{i}(a_{11} + a_{22}) \lambda}
    d\lambda \int_\mathbb{R} \mathrm{e}^{- \varepsilon \eta^2 -
    \mathrm{i}(a_{11}-a_{22}-2a_{12}(1-2\varepsilon))\eta}d\eta\;,
\end{eqnarray*}
where the right-hand side has been reduced to a product of Riemann
integrals by Fubini's theorem and the definition of what it means to
integrate a differential form.

The crucial point now is that the integral over $\eta$ vanishes in
the limit $\varepsilon \to 0\,$. Indeed, writing $b_\varepsilon :=
\frac{1}{2}(a_{11} - a_{22}) - a_{12} (1-2\varepsilon)$ for short,
the $\eta$-integral gives
\begin{displaymath}
    \int_{\mathbb{R}} \mathrm{e}^{- \varepsilon \eta^2 -
    2\mathrm{i}b_\varepsilon \eta} d\eta = \mathrm{e}^{-
    b_\varepsilon^2 / \varepsilon} \sqrt{\pi / \varepsilon} \;,
\end{displaymath}
and this does go to zero for $\varepsilon \to 0$ as long as the real
number $b_0$ is non-zero. The latter condition is always satisfied,
as the constraints on the matrix elements of $A$ ensure that
\begin{displaymath}
    b_0 = {\textstyle{\frac{1}{2}}} (a_{11} - a_{22}) - a_{12}
    \ge \sqrt{-a_{11} a_{22}} - |a_{12}| > 0 \;.
\end{displaymath}
Thus we have shown that $\lim_{\varepsilon \to 0+} \int_E \psi^\ast
\mathrm{i} G_{A,\varepsilon}\, \iota_{\tau(\dot{A})} |dR| = 0\,$.

The situation is no different for the integral over the 2-plane $F$.
We therefore conclude that $\frac{d}{dt} I(A + t\dot{A}) \vert_{t =
0} = 0\,$. Hence $I(A) =: C_{1,1}^{-1}$ is a constant independent of
$A$ for $A s > 0\,$. This proves the validity of the
Hubbard-Stratonovich transformation in the following precise form
with cutoff function $\chi_\varepsilon (R) = \mathrm{e}^{-
\frac{\varepsilon}{2} \mathrm{Tr}\,(sR - Rs)^2}$:
\begin{displaymath}
    \mathrm{e}^{-\mathrm{Tr}\,A^2} = C_{1,1}\lim_{\varepsilon\to 0+}
    \sum_{\sigma\in \{\bullet\circ,\circ\bullet\}} \mathrm{sgn}(\sigma)
    \int_{D_\sigma}\mathrm{e}^{-\mathrm{Tr}\,R^2 - 2\mathrm{i}\,
    \mathrm{Tr}\, AR} \chi_\varepsilon(R) |dR| \;,
\end{displaymath}
where $\mathrm{sgn}(\bullet\circ)= 1$ and $\mathrm{sgn}(\circ\bullet)
= -1$, and we have replaced $\chi_\varepsilon(R+\mathrm{i}A)$ by
$\chi_\varepsilon(R)$.

Let us emphasize once again that the right-hand side is \emph{not} a
standard Gaussian integral, and that our proof of this formula was
not by completing the square and shifting. We will show in the next
subsection that $C_{1,1} = \mathrm{i}\, 2^{\frac{1}{2}} \pi^{-
\frac{3}{2}}$.

\subsection{Short proof}

We now spell out another line of approach for the case of $p = q =
1$, which is a variant of that given in \cite{yf1}. This calculation
will be much quicker, but does not extend (at least not in any way
known to us) to higher values of $p,\,q\,$.

Recall that in light-cone coordinates $\xi$, $\eta$, and $\lambda$,
we have the expression
\begin{displaymath}
    \mathrm{e}^{-\frac{1}{2} \mathrm{Tr}\, R^2 - \mathrm{i}\,
    \mathrm{Tr}\, A R } = \mathrm{e}^{-\lambda^2 - \mathrm{i}(a_{11}
    + a_{22})\lambda}\, \mathrm{e}^{- \xi\eta - \mathrm{i} b_0 \eta
    - \mathrm{i} b_1 \xi}
\end{displaymath}
where $b_0 = \frac{1}{2} (a_{11} - a_{22}) - a_{12} > 0$ and $b_1 =
\frac{1}{2} (a_{11} - a_{22}) + a_{12} > 0\,$. It is clear that this
can be integrated against $\int_\mathbb{R} d\lambda$ resulting in a
factor $\sqrt{\pi}\, \mathrm{e}^{- \frac{1}{4} (a_{11} + a_{22})^2}$.
It then remains to integrate the other factor $\mathrm{e}^{- \xi\eta
- \mathrm{i} b_0 \eta - \mathrm{i} b_1 \xi}$ against $d\xi \,d\eta$
over the two quadrants $\xi,\eta > 0$ and $\xi,\eta < 0\,$. For this
we make the substitution of variables $\eta = \rho\,\mathrm{e}^\tau$
and $\xi = \rho\,\mathrm{e}^{-\tau}$ which transforms the measure
$d\xi \, d\eta$ to $2 | \rho|\, d\rho d\tau\,$. Noticing that $\rho
> 0$ on $D_{\bullet\circ}\,$, $\rho < 0$ on $D_{\circ\bullet}\,$,
and our chain of integration is $D = D_{\bullet\circ} - D_{\circ
\bullet}\,$, we drop the absolute value on $|\rho|$ and integrate
against the sign-alternating distribution $\int_{\mathbb{R}^2} \rho
d\rho d\tau\,$. More precisely, we take the $\rho$ integration to be
the inner one (this obviates the need for regularization albeit at
the expense of Fubini's theorem becoming inapplicable, so that the
order of integrations is now fixed and can no longer be changed). The
resulting $\rho$ integral is
\begin{displaymath}
    \int_\mathbb{R} \mathrm{e}^{-\rho^2 - 2 \mathrm{i} \rho\beta}
    \rho \, d\rho = - \mathrm{i} \sqrt{\pi}\, \beta\, \mathrm{e}^{
    -\beta^2} \;, \qquad \beta = {\textstyle{\frac{1}{2}}}
    (b_0\, \mathrm{e}^\tau + b_1\, \mathrm{e}^{ -\tau}) \;.
\end{displaymath}
We now observe the relation
\begin{displaymath}
    \beta\, \mathrm{e}^{-\beta^2} d\tau = \mathrm{e}^{-b_0 b_1}
    \, \mathrm{e}^{-\frac{1}{4} (b_0\, \mathrm{e}^\tau - b_1
    \mathrm{e}^{-\tau})^2}{\textstyle{\frac{1}{2}}}\, d \left(
    b_0\, \mathrm{e}^\tau - b_1\, \mathrm{e}^{-\tau} \right)\;,
\end{displaymath}
which shows that the remaining $\tau$ integral is another standard
Gaussian integral in the variable $\frac{1}{2}(b_0\, \mathrm{e}^\tau
- b_1\, \mathrm{e}^{-\tau})$, yielding the value $\sqrt{\pi}\,
\mathrm{e}^{-b_0 b_1}$. Altogether we then obtain
\begin{displaymath}
    C_{1,1} \int_D \mathrm{e}^{-\frac{1}{2}\mathrm{Tr}\,R^2 -
    \mathrm{i} \mathrm{Tr}\, AR} \, |dR| = 2^{\frac{3}{2}}
    \mathrm{e}^{-\frac{1}{2}\mathrm{Tr}\,A^2}\;,\quad C_{1,1}
    = \mathrm{i}\, 2^{\frac{1}{2}} \pi^{-\frac{3}{2}} \;,
\end{displaymath}
for $D = D_{\bullet\circ} - D_{\circ\bullet}\,$, which is the desired
result but for scaling $A \to \sqrt{2} \, A$ and $R \to \sqrt{2}\,
R\,$. (Of course it must be understood here that the $\rho$
integration has to be done first.)

This computation, while direct and short, does not seem to carry over
to the case of higher values of $p$ and $q\,$, whereas our conceptual
proof of Sects.\ \ref{sect:4.1}-\ref{sect:4.3} does.

\section{The general case of $\mathrm{O}_{p,\,q}$-symmetry}
\label{sect:p,q}\setcounter{equation}{0}

We now turn to the general case of real matrices $R$ of size $(p+q)
\times (p+q)$,
\begin{displaymath}
    R = \begin{pmatrix} R_{p,p} &R_{p,\,q} \\
    -(R_{p,\,q})^\mathrm{t} &R_{q,\,q} \end{pmatrix} \;,
\end{displaymath}
where the blocks $R_{p,p}$ and $R_{q,\,q}$ are symmetric. Such
matrices $R$ as a whole satisfy the linear symmetry relation
\begin{displaymath}
    R = s R^\mathrm{t} s \;, \quad s =
    \mathrm{diag}( \mathrm{Id}_p \;, - \mathrm{Id}_q ) \;.
\end{displaymath}
Equivalently, if we equip the real vector space $\mathbb{R}^{p+q}$
with the indefinite bilinear form
\begin{displaymath}
    B(u,v) \equiv u^\mathrm{t} s v := \sum_{i=1}^p u_i \, v_i
    - \sum_{j=p+1}^{p+q} u_j \, v_j \;,
\end{displaymath}
then our matrices $R$ form the linear space $\mathrm{Sym}_B(
\mathbb{R}^{p+q})$ of $B$-symmetric matrices, i.e.,
\begin{displaymath}
    \forall \, u \, , v \in \mathbb{R}^{p+q} : \quad B(u,R\,v) =
    B(v, R\,u)\;.
\end{displaymath}
The symmetry group of the bilinear form $B$ with signature $(p,q)$ is
the non-compact real orthogonal group $\mathrm{O}_{p,\,q}$ of
matrices $g$ satisfying the equation $s = g^\mathrm{t} s g$ or
equivalently, $B(u,v) = B(gu,gv)$ for all $u\, , v \in
\mathbb{R}^{p+q}$.

We still want our matrices $R \in \mathrm{Sym}_B(\mathbb{R}^{p+q})$
to be diagonalizable by the $\mathrm{O}_{p,\,q}$-action $R \mapsto g
R\, g^{-1}$. In the simple case $p = q = 1$ we saw that
diagonalizability fails when the off-diagonal block $R_{p,\,q}$ is
larger than the difference of the diagonal blocks $R_{p,p}$ and
$R_{q,\,q}\,$. A similar phenomenon is expected to occur for general
$p , q\,$. Roughly speaking, $R$ will fail to have real eigenvectors
and eigenvalues when the degree of non-Hermiticity coming from the
off-diagonal blocks becomes too large.

\subsection{Description of components $D_\sigma\,$}
\label{sect:2.3.1}

In the Hubbard-Stratonovich integral (\ref{eq:sum1}) we integrate
over the set, $D$, of $R \in \mathrm{Sym}_B(\mathbb{R}^{p+q})$ that
can be brought to diagonal form by the $\mathrm{O}_{p, \,q}$-action,
i.e., matrices of the form $R = g \lambda g^{-1}$ with $\lambda =
\mathrm{diag}(\lambda_1,\lambda_2,\ldots,\lambda_{p+q})$ and $g \in
\mathrm{O}_{p,\,q}\,$. We know from our considerations for $p = q =
1$ that $D$ decomposes into connected sets $D_\sigma$ whose
boundaries intersect in lower-dimensional sets (of codimension 2).
Here we describe these connected components $D_\sigma\,$.

Let $R = g \lambda g^{-1}$ be $\mathrm{O}_{p,\,q}$-diagonalizable in
the sense just described, so that the $p+q$ column vectors which make
up $g$ form a basis of $\mathbb{R}^{p+q}$ consisting of eigenvectors
of the matrix $R\,$. Let $v^{(k)}$ denote the $k^\mathrm{th}$ column
of $g$. Then by the equation $g^\mathrm{t} s g = s$ the vector $v^{
(k)}$ has positive norm square $B(v^{(k)},v^{(k)}) = 1 > 0$ for $k =
1, \ldots, p$ and negative norm square $B(v^{(k)},v^{(k)}) = -1 < 0$
for $k = p+1, \ldots, p+q\,$. To distinguish between these two cases,
we use the language of relativity theory and call the corresponding
eigenvalue $\lambda_k$ of `space-like' type in the former case and
'time-like' in the latter.

Assuming that we are in the generic situation, where all eigenvalues
$\lambda_k$ of $R$ differ from each other, we can arrange them as a
descending sequence. Given any such sequence, we assign to it a
motif, $\sigma$: if an eigenvalue of the ordered sequence is
space-like, then we represent it as a full dot, otherwise as an empty
dot. Thus our motifs $\sigma$ are sequences of $p$ full and $q$ empty
dots; e.g., $\sigma = \bullet \circ \circ \bullet \circ \bullet$ is a
motif for $p = q = 3$.

With each motif $\sigma$ we now associate a domain $D_\sigma$ as
follows. While insisting that eigenvalues of opposite type
(space-like vs.\ time-like) must never be equal, we lift the
condition of non-degeneracy of eigenvalues partially: eigenvalues of
the same type are now allowed to collide and exchange positions, but
eigenvalues of opposite type still are not. This rule of eigenvalue
motion defines a connected domain $D_\sigma$ for each $\sigma\,$.

We define a sign function $\sigma \mapsto \mathrm{sgn}(\sigma) \in \{
\pm 1 \}$ by setting $\mathrm{sgn}(\sigma) := (-1)^n$ where $n$ is
the number of transpositions $\bullet\circ \leftrightarrow
\circ\bullet$ which are needed to reduce $\sigma$ to the reference
motif $\sigma_0 = \bullet \bullet \cdots \bullet \circ \circ \cdots
\circ$ made of $q$ empty dots following $p$ full dots. The parity of
this number $n$ is well-defined although $n$ itself is not. Indeed,
another expression for $\mathrm{sgn}(\sigma) = (-1)^n$ is
\begin{displaymath}
    \mathrm{sgn}(\sigma) = \prod_{i=1}^p \prod_{j=p+1}^{p+1}
    \mathrm{sign}(\lambda_i - \lambda_j) \;,
\end{displaymath}
where the $\lambda$'s are the eigenvalues of any $R \in D_\sigma\,$.

\subsection{Description of boundary}

As was explained above, a point in the interior of $D_\sigma$ is a
matrix $R\in\mathrm{Sym}_B(\mathbb{R}^{p+q})$ which is diagonalizable
by the $\mathrm{O}_{p,\,q}$-action and has eigenvalues that order
according to the motif $\sigma$ -- in particular, no two eigenvalues
of opposite type are equal. This means that if $t \mapsto R(t)$ is a
continuous curve which starts in $D_\sigma$ but leads to a collision
of two or more eigenvalues of opposite type at $t_c\,$, then $R(t_c)$
is a point in the boundary of $D_\sigma\,$. The generic boundary
situation is that (i) exactly two eigenvalues of opposite type
collide, (ii) the norms of the two corresponding eigenvectors go to
zero, and (iii) the boundary point $R(t_c)$ is a matrix which fails
to be diagonalizable by any real transformation $g \in
\mathrm{O}_{p,\,q} \,$.

To carry out the computations leading to the desired result
(\ref{eq:indep}), we need a good description of the interior of $D =
\bigcup_\sigma D_\sigma$ as well as the boundary $\partial D$. This
is achieved as follows. Let us first make a slight change of
perspective: abandoning the viewpoint of $R$ as a matrix, we switch
to regarding $R$ as an invariantly defined linear operator (giving
rise to a matrix when expressed w.r.t.\ some fixed basis); i.e.,
taking the matrix elements $R^{ij}$ we let $R := R^{ij} E_{ij}$
(summation convention!) with $E_{ij}$ the standard basis of $\mathrm
{End}(\mathbb{R}^{p+q})$ defined in terms of the standard basis $e_i$
of $\mathbb{R}^{p+q}$ by $E_{ij}\, e_k = e_i\, \delta_{jk}\,$.

Consider now the Grassmannian $\mathrm{Gr}_{1,1} (\mathbb{R}^{p+q})$
of $B$-orthogonal decompositions
\begin{displaymath}
    \mathbb{R}^{p+q} = V \oplus V^\perp \;,
\end{displaymath}
where $V \subset \mathbb{R}^{p+q}$ is any Lorentzian plane, i.e., a
$2$-dimensional subspace $V = L^+ \oplus L^-$ spanned by a pair of
null lines or light-like directions $L^\pm$ such that $s(L^\pm) =
L^\mp$. The restriction of $B$ to $V$ is then non-degenerate, and the
$B$-orthogonal complement $V^\perp$ of $V$ is spanned by $p-1$
space-like and $q-1$ time-like vectors. Note that $\mathrm{Gr}_{1,1}
(\mathbb{R}^{p+q})$ is acted upon transitively by the group $\mathrm
{SO}_{p,\,q}\,$, which is to say that every Lorentzian plane $V
\subset \mathbb{R}^{p+q}$ can be regarded as the image $g(\mathbb
{R}^{1+1})$ of the standard Lorentzian plane $\mathbb{R}^{ 1+1}$ by
some transformation $g \in \mathrm{SO}_{p,\, q}\,$. Since
$\mathbb{R}^{1+1}$ has isotropy $\mathrm{S}(\mathrm{O}_{1,1} \times
\mathrm{O}_{p-1,\,q-1})$, our manifold $\mathrm{Gr}_{1,1}
(\mathbb{R}^{p+q})$ is the base of a principal fibre bundle
\begin{equation}\label{eq:PFB}
    \mathrm{SO}_{p,\,q} \to \mathrm{SO}_{p,\,q} / \mathrm{S}
    (\mathrm{O}_{1,1} \times \mathrm{O}_{p-1,\, q-1}) \simeq
    \mathrm{Gr}_{1,1} (\mathbb{R}^{p+q}) \;.
\end{equation}
Moreover, since the Lorentzian planes $V \subset \mathbb{R}^{p+q}$
are in one-to-one correspondence with pairs of real lines, one in
$\mathbb{R}^p$ and $\mathbb{R}^q$ each, $\mathrm{Gr}_{1,1}
(\mathbb{R}^{p+q})$ is a product
\begin{displaymath}
    \mathrm{Gr}_{1,1}(\mathbb{R}^{p+q}) \simeq
    \mathbb{R}\mathrm{P}_{p-1} \times \mathbb{R}\mathrm{P}_{q-1}
\end{displaymath}
of real projective spaces $\mathbb{R}\mathrm{P}_{n-1} \equiv
\mathbb{R}^n / (\mathbb{R} \setminus \{0\})$ for $n = p$ and $n =
q\,$. From this identification it follows immediately that the
manifold $\mathrm{Gr}_{1,1} (\mathbb{R}^{p+q})$ is compact,
non-simply connected, and non-orientable unless both $p$ and $q$ are
even. It is also seen that (\ref{eq:PFB}) decomposes as a direct
product of two principal fibre bundles
\begin{displaymath}
    \mathrm{SO}_n \to \mathrm{SO}_n / \mathrm{S} (\mathrm{O}_1
    \times \mathrm{O}_{n-1}) \simeq \mathbb{R} \mathrm{P}_{n-1}
\end{displaymath}
for $n = p$, $q\,$. We note that these principal fibre bundles are
non-trivial.

Choosing a point $x \in \mathrm{Gr}_{1,1}( \mathbb{R}^{p+q} )$ is
equivalent to choosing an orthogonal projector
\begin{displaymath}
    \Pi_x \, : \,\, \mathbb{R}^{p+q} \to V_x \equiv V \;.
\end{displaymath}
By orthogonality, $B$ restricts to a bilinear form of signature
$(1,1)$ on $V_x$ and a bilinear form of signature $(p-1,q-1)$ on the
orthogonal complement $V_x^\perp \equiv (\mathrm{Id}-\Pi_x)
\mathbb{R}^{p+ q}$. We denote by $\mathrm{O}(V_x)$ resp.\ $\mathrm{O}
(V_x^\perp)$ the symmetry groups of these symmetric bilinear forms.
Let $\mathrm {Sym}_B (V_x)$ denote the subspace of linear operators
$r:\, V_x \to V_x$ which are symmetric with respect to $B$ restricted
to $V_x\,$, and let $D(V_x^\perp) \subset \mathrm{Sym}_B( V_x^\perp)$
denote the subspace of $\mathrm{O}( V_x^\perp)$-diagonalizable linear
transformations $t : \, V_x^\perp \to V_x^\perp\,$.

Now consider the set, $\mathcal{X}$, of triples $(x\,;r\,,t)$
consisting of any point $x \in \mathrm{Gr}_{1,1}( \mathbb{R}^{p+q} )$
and two linear operators $r \in \mathrm{Sym}_B(V_x)$ and $t \in
D(V_x^\perp)$. This set $\mathcal{X}$ has the structure of a fibre
bundle
\begin{displaymath}
    \pi : \, \mathcal{X} \to \mathrm{Gr}_{1,1}(\mathbb{R}^{p+q})
    \;, \quad (x\,; r \, , t) \mapsto x \;,
\end{displaymath}
where the fibre over $x$ is a direct product
\begin{displaymath}
    \pi^{-1}(x) = \mathrm{Sym}_B(V_x) \times D(V_x^\perp) \;.
\end{displaymath}
One may also view $\mathcal{X}$ as an associated bundle $\mathcal{X}
= G \times_K W$ with $G = \mathrm{SO}_p \times \mathrm{SO}_q\,$,
\begin{displaymath}
    K = \mathrm{S}(\mathrm{O}_1 \times \mathrm{O}_{p-1}) \times
    \mathrm{S}(\mathrm{O}_1 \times \mathrm{O}_{q-1}) \;, \quad
    W = \mathrm{Sym}_B(\mathbb{R}^{1+1}) \times D(\mathbb{R}^{
    (p-1) + (q-1)}) \;,
\end{displaymath}
where the action of $K$ on $W$ is by conjugation. Since the principal
bundle $G \to G/K = \mathrm{Gr}_{1,1}( \mathbb{R}^{p+q})$ is
non-trivial, so is the associated bundle $\mathcal{X} \to
\mathrm{Gr}_{1,1}(\mathbb{R}^{p+q})$.

$\mathcal{X}$ can be viewed as a subset of our space $\mathrm{Sym}
_B(\mathbb{R}^{p+q})$ of operators $R\,$. Indeed, for $R \equiv (x\,;
r , t)$ we may use the projector $\Pi_x$ to define $R\, v$ for any
vector $v \in \mathbb{R}^{p+q}$ by
\begin{displaymath}
    R\, v := r\, \Pi_x\, v + t\, (\mathrm{Id} - \Pi_x ) v \;.
\end{displaymath}
In other words, $R$ acts on $V_x$ as $r$ and on $V_x^\perp$ as $t\,$.
Since both $r$ and $t$ are symmetric with respect to $B$, so is $R =
r \, \Pi_x + t \, (\mathrm{Id} - \Pi_x)$. More formally, this
correspondence $(x\,; r , t) \mapsto r\, \Pi_x + t\, (\mathrm{Id} -
\Pi_x )$ defines a mapping $\phi : \, \mathcal{X} \to \mathrm{Sym}_B
( \mathbb{R}^{p+q})$. Note that the mapping $\phi$ is not surjective,
as the operators $R$ with more than two non-real eigenvalues cannot
be presented in this form. However, the image $\phi(\mathcal{X})$ is
large enough to contain our integration domain $D$ and also (in the
measure-theoretic sense) the boundary $\partial D$.

Now for any point $x$ of the Grassmann manifold $\mathrm{Gr}_{1,1}
(\mathbb{R}^{p+q})$ let $L_x^\pm$ denote the two null lines of the
Lorentzian vector space $V_x\,$. In other words $B(u,u) = 0$ for $u
\in L_x^+$ or $u \in L_x^-\,$. The three-dimensional space $\mathrm
{Sym}_B (V_x)$ then decomposes as
\begin{equation}\label{eq:inv-decomp}
    \mathrm{Sym}_B(V_x) = \mathbb{R}\, \mathrm{Id}_{V_x} \oplus
    \mathrm{Hom}(L_x^+,L_x^-) \oplus \mathrm{Hom}(L_x^-,L_x^+)\;.
\end{equation}
Let us choose a basis $e_x^\pm \in L_x^\pm$ of null vectors of $V_x$
so that $s e_x^\pm = e_x^\mp$ and $B(e_x^+ , e_x^-) = 1$. In such a
basis $\Pi_x$ is expressed as $\Pi_x = e_x^+ B(e_x^- , \cdot) + e_x^-
B(e_x^+ ,\cdot)$. Linear coordinates $\lambda$, $\xi$, $\eta$ for
$\mathrm{Sym}_B (V_x)$ are then introduced by decomposing $r$
according to (\ref{eq:inv-decomp}):
\begin{displaymath}
    r = \lambda\, \Pi_x + \xi\, e_x^+ B(e_x^+,\cdot)
    + \eta\, e_x^- B(e_x^-,\cdot) \;.
\end{displaymath}

Extending the notation used before, let $E_x \in \mathrm{Sym}_B(V_x)$
and $F_x \subset \mathrm{Sym}_B(V_x)$ denote the linear subspaces
which are defined by the equations $\xi = 0$ and $\eta = 0$
respectively. These 2-planes $E_x$ and $F_x$ will assume the roles
played by $E$ and $F$ in Sect.\ \ref{sect:4.2}. It should be
emphasized, however, that $\xi$ and $\eta$ do not extend to globally
defined coordinate functions for $\mathcal {X}$. The point here is
that transporting the pair of null lines $L_x^\pm$ around a closed
path in $\mathrm{Gr}_{1,1}(\mathbb{R} ^{p+q})$ one may find that
$L_x^+$ and $L_x^-$ get interchanged. In other words, although
$L_x^\pm$ are invariantly defined as a \emph{pair} of lines, there is
no invariant way of telling which line is $L_x^+$ and which is
$L_x^-$. As a result, the linear coordinate functions $\xi$ and
$\eta$ projecting from $\mathrm{Sym}_B(V_x)$ to $\mathrm{Hom} (L_x^-
, L_x^+)$ and $\mathrm{Hom}(L_x^+ , L_x^-)$ are only defined locally
in the variable $x \in \mathrm{Gr}_{1,1}(\mathbb{R}^{p+q})$. By the
same token, the union of planes $E_x \cup F_x$ is invariantly defined
for all $x\,$, whereas the planes $E_x$ and $F_x$ individually are
not.

The smooth assignment $\mathrm{Sym}_B(V_x) \supset E_x \cup F_x
\mapsto x$ gives a bundle $\mathcal{E} \subset \mathcal{X}$ with
fibre $(E_x \cup F_x) \times D(V_x^\perp)$ over $x \in
\mathrm{Gr}_{1,1} (\mathbb{R}^{p+q})$. By applying the map $\phi : \,
\mathcal{X} \to \mathrm{Sym}_B(\mathbb{R}^{p+q})$ we then obtain a
submanifold $\phi(\mathcal{E}) \subset \partial D$. As a matter of
fact, since the generic situation at the boundary of $D =
D(\mathbb{R}^{p+q})$ is that exactly two eigenvalues of opposite type
collide, the set $\phi(\mathcal{E})$ agrees with the set $\partial D$
up to lower-dimensional pieces.

\subsection{Orientation of boundary}

We now introduce an outer orientation on $\mathcal{E}$ to achieve the
stronger property that $\phi(\mathcal{E})$ agrees with the boundary
$\partial D$ as a chain (or distribution). For that, fix any $x \in
\mathrm{Gr}_{1,1}(\mathbb{R}^{p+q})$ and consider some $(r,t)\in E_x
\times D(V_x^\perp)$. Arrange the single eigenvalue $\lambda(r)$ of
$r$ and the eigenvalues $\lambda_1,\lambda_2, \ldots, \lambda_{p+q
-2}$ of $t$ as a decreasing sequence. (This ordering is uniquely
defined except for the set of measure zero where some eigenvalues are
degenerate.) Following the prescription of Sect.\ \ref{sect:2.3.1},
draw the space-like eigenvalues of $t$ as full dots and the time-like
eigenvalues as empty dots. In view of Fig.\ \ref{fig:Or}, if $\eta(r)
> 0$ then represent $\lambda(r)$ by $\bullet \circ$, else by $\circ
\bullet$. This defines a motif $\sigma(r,t)$ for almost every pair
$(r,t)$. For $p = q = 2$ for example, if $\lambda_1 > \lambda_2
> \lambda(r)$, $\lambda_1$ is space-like, $\lambda_2$ is time-like,
and $\eta(r) > 0$, then $\sigma(r,t) = \bullet\circ\bullet \circ$.

The point $\phi(x\,;r,t)$ is a point in the boundary of $D_{\sigma(r,
\,t)} \subset \partial D$. According to the alternating sign of the
sum $\partial D = \sum \mathrm{sgn}(\sigma) \partial D_\sigma$ the
outer orientation of $\partial D$ at $\phi(x\,;r,t)$ is directed from
the inside to the outside if $\mathrm{sgn}(\sigma(r,t)) = +1$ and
from the outside to the inside if $\mathrm{sgn}(\sigma(r,t)) = -1$.
Consulting again Fig.\ \ref{fig:Or} we assign to the boundary point
$\phi(x\,;r,t)$ the transverse vector
\begin{displaymath}
    - \mathrm{sign}(\eta(r)) \, \mathrm{sgn}(\sigma(r,t))
    \, (\phi_\ast)_{(x\,;\,r,\,t)}(\partial_\xi) \;,
\end{displaymath}
where $(\phi_\ast)_{x\,;r,\,t}$ means the differential $d\phi \equiv
\phi_\ast$ of the map $\phi$ at the point $(x\,;r,t)$. The motif
$\sigma(r,t)$ switches sign as $\eta(r)$ passes through zero;
therefore our assignment is smooth along the $\eta$-axis of $E_x\,$,
although discontinuities may occur when the eigenvalue $\lambda(r)$
hits eigenvalues of $t$ or when eigenvalues of $t$ of opposite type
collide.

In the case of the other component, $F_x \times D(V_x^\perp)$, we
proceed in the same way, albeit with the two boundary coordinate
functions $\xi$ and $\eta$ exchanging roles. Thus we assign to a
point $\phi(x\,;r,t)$ for $(r,t) \in F_x \times D(V_x^\perp)$ the
transverse vector
\begin{displaymath}
    - \mathrm{sign}(\xi(r)) \, \mathrm{sgn}(\sigma(r,t)) \,
    (\phi_\ast)_{(x\,;\, r,\,t)}(\partial_\eta) \;,
\end{displaymath}
which is again smooth as a function of $\xi$.

Now while the distinction between $\xi$ and $\eta$ (or $E_x$ and
$F_x$) is only defined locally in $x$, the outcome of our discussion
is free of the ambiguity of exchanging $\xi \leftrightarrow \eta$.
Indeed, a glance at Fig.\ \ref{fig:Or} shows that the picture is
invariant under the reflection ($\xi \leftrightarrow \eta$) at the
axis dividing the first quadrant. Hence our procedure gives a
\emph{globally} defined piecewise smooth transverse vector field and
hence a piecewise smooth outer orientation of the boundary
$\phi(\mathcal{E})$. Adopting this outer orientation we have
\begin{displaymath}
    \phi(\mathcal{E}) = \sum\nolimits_\sigma
    \mathrm{sgn}(\sigma) \partial D_\sigma
\end{displaymath}
as an equality between chains, i.e., in the weak sense:
\begin{equation}\label{eq:bound-final}
    \int_{\phi(\mathcal{E})} \omega = \sum_\sigma
    \mathrm{sgn}(\sigma) \int_{D_\sigma} d\omega \;,
\end{equation}
for any bounded twisted differential form $d\omega$ of top degree.
This result holds true, in particular, for the integrand $\omega =
\mathrm{i} G_{A,\varepsilon} \, \iota_{\tau (\dot{A})} \vert dR
\vert$ of equation (\ref{eq:dIda}) with the Gaussian
\begin{displaymath}
    G_{A,\varepsilon}(R) = \mathrm{e}^{-\mathrm{Tr}\,
    (R + \mathrm{i} A)^2} \chi_\varepsilon(R + \mathrm{i}A)
\end{displaymath}
regularized by the cutoff function $\chi_\varepsilon(R) =
\mathrm{e}^{- \frac{\varepsilon}{2} \mathrm{Tr}\,(sR - Rs)^2}$.

\subsection{Pulling back to $\mathcal{E}\,$}

In the next step the left-hand side of the integral
(\ref{eq:bound-final}) for $\omega = \mathrm{i}G_{A,\varepsilon}\,
\iota_{\tau(\dot{A})} |dR|$ is pulled back to $\mathcal{E} \subset
\mathcal{X}$. This results in an iterated integral where the outer
integral is over the compact manifold $\mathrm{Gr}_{1,1}(\mathbb{R}
^{p+q}) \ni x$ and the inner integral is over $(E_x \cup F_x) \times
D(V_x^\perp)$. It will now be shown that the inner integral over $E_x
\cup F_x$ is of exactly the same form as in the case of $p = q = 1$
and hence, for the reasons explained in Sect.\ \ref{sect:4.3},
vanishes in the limit $\varepsilon \to 0\,$. We will do this by
inspection of $\phi^\ast \omega$.

First we compute the pull back $\phi^\ast(\iota_{\tau(\dot{A})}
|dR|)$. Denoting by $(\phi^{-1})_\ast = (\phi_\ast)^{-1} \equiv
\phi_\ast^{ -1}$ the differential of the inverse map $\phi^{-1}$ we
have
\begin{displaymath}
    \phi^\ast( \iota_{ \tau(\dot{A})} \vert dR \vert) = \iota_{
    \phi_\ast^{-1} (\tau(\dot{A}))} \phi^\ast \vert dR \vert \;.
\end{displaymath}
With this relation in mind we now compute $\phi^\ast |dR|$. For this
we notice that the map $\phi :\, \mathcal{X}\to \mathrm{Sym}_B
(\mathbb{R}^{p+q})$, $(x\,;r,t)\mapsto r \,\Pi_x + t (\mathrm{Id}-
\Pi_x)$ schematically has the variation
\begin{displaymath}
    \delta \phi = (\delta r) \Pi_x + (\delta t) (\mathrm{Id} - \Pi_x)
    + (r-t) \delta \Pi_x \;.
\end{displaymath}
The first two terms contribute factors of unity to the Jacobian when
one pulls back the translation-invariant measure $|dR|$ of $\mathrm
{Sym}_B ( \mathbb{R}^{p + q})$.

Now the third term is purely off-diagonal w.r.t.\ the decomposition
$\mathbb{R}^{p+q} = V_x^{\vphantom{\perp}} \oplus V_x^\perp :$
\begin{displaymath}
    \delta \Pi_x \in \big( \mathrm{Hom}(V_x^{\vphantom{\perp}}\,,
    V_x^\perp)\oplus\mathrm{Hom}(V_x^\perp , V_x^{\vphantom{\perp}}
    ) \big) \cap \mathrm{Sym}_B( \mathbb{R}^{p+q} )\;,
\end{displaymath}
and may be expressed as a commutator $\delta \Pi_x = [\alpha, \Pi_x]$
with a generator $\alpha \in \mathrm{Lie}(\mathrm{O}_{p,\,q})$. The
two components of $\delta \Pi_x$ in $\mathrm{Hom}(V_x^{\vphantom{
\perp}}\, ,V_x^\perp)$ and $\mathrm{Hom}(V_x^\perp, V_x^{\vphantom{
\perp}})$ are related to each other by the condition of $B$-symmetry.
Identifying $V_x$ with its dual $V_x^\ast$ by means of the
non-degenerate form $B \vert_{V_x \times V_x}$ we have an isomorphism
$\mathrm{Hom} (V_x^{ \vphantom{\perp}}\, ,V_x^\perp) \simeq
V_x^{\vphantom{\perp}} \otimes V_x^\perp\,$. By this isomorphism we
may view the multiplication operator $\delta \Pi_x \mapsto (r-t)
\delta \Pi_x$ as a linear transformation $r \otimes \mathrm{Id} -
\mathrm{Id} \otimes t$ on $V_x^{\vphantom{ \perp}} \otimes
V_x^\perp$. We thus arrive at the formula
\begin{displaymath}
    \phi_{x\,; r_x\, ,\, t_x}^\ast |dR| =
    J_x(r_x\,,t_x)\, dx \; dr_x \; dt_x \;,
\end{displaymath}
with the Jacobian of the transformation being
\begin{displaymath}
    J_x(a,b) = \big\vert \mathrm{Det}(a \otimes \mathrm{Id}_
    {V_x^\perp} - \mathrm{Id}_{V_x^{\vphantom{\perp}}}\otimes b)
    \big\vert \;.
\end{displaymath}
Here $dx$ denotes a suitably normalized $\mathrm{SO}_{p,\,q}
$-invariant measure for the compact manifold $\mathrm{Gr}_{1,1}
(\mathbb{R}^{p+q})$, while $dr_x$ and $dt_x$ are
translation-invariant measures for $\mathrm{Sym}_B(V_x)$ and
$D(V_x^\perp)$. As an observation to be used presently, we note that
the Jacobian
\begin{displaymath}
    J_x(a,b) = \prod_{k=1}^{p+q-2} \left\vert (\lambda(a) -
    \lambda_k(b))^2 - \xi(a) \eta(a) \right\vert
\end{displaymath}
becomes independent of $\xi$ and $\eta$ for $a \in E_x \cup F_x\,$.

We now have to contract the density $\phi^\ast |dR|$ with the vector
field $\phi_\ast^{-1}( \tau(\dot{A}) )$ and make the restriction to
$\mathcal{E}$. Since $d\xi = 0$ on $E_x$ and $d\eta = 0$ on $F_x\,$,
the contraction after restriction depends only on the $\xi$- and
$\eta$-components of $\phi_\ast^{-1}( \tau( \dot{A}) )$:
\begin{displaymath}
    \phi_\ast^{-1}( \tau(\dot{A}) ) = \dot{A}^\xi \partial_\xi +
    \dot{A}^\eta \partial_\eta + \ldots \;.
\end{displaymath}
While these components vary with $x\,$,
\begin{displaymath}
    \dot{A}^\xi(x) = B(e_x^- , \dot{A} e_x^-) \;, \quad
    \dot{A}^\eta(x) = B(e_x^+ , \dot{A} e_x^+) \;,
\end{displaymath}
they are constant along $E_x$ and $F_x\,$. Hence, recalling the
expression $dr_x = d\lambda \, d\xi \, d\eta$ and assembling factors,
we see that $\phi^\ast (\iota_{\tau(\dot{A})} |dR|)$ for fixed $x$ is
proportional to the form $\pm d\lambda \, d\eta$ (with constant
coefficient) along $E_x$ and $\pm d\lambda \, d\xi$ (still with
constant coefficient) along $F_x\,$. In other words, the inner
dependence (along $E_x$ and $F_x$) of the integration form
$\phi^\ast( \iota_{\tau(\dot{A})} |dR|)$ is the same as in Sect.\
\ref{sect:4.3}.

\subsection{Gaussian integrand on boundary}

Our final step is to compute the restriction to $\mathcal{E}$ of the
pull back $\phi^\ast G_{A, \varepsilon}$ of the Gaussian $G_{A,
\varepsilon}\,$. For this we decompose
\begin{displaymath}
    \mathrm{Tr}\, (R + \mathrm{i}A)^2 = \mathrm{Tr}\, R^2 +
    2\mathrm{i}\, \mathrm{Tr}\, A R - \mathrm{Tr}\, A^2 =:
    f_2(R) + f_1(R) + f_0 \;.
\end{displaymath}
Fixing any $x \in \mathrm{Gr}_{1,1}(\mathbb{R}^{p+q})$ and inserting
$R = \phi(x\,;r,t) = r\,\Pi_x + t\,(\mathrm{Id} - \Pi_x)$ we obtain
\begin{eqnarray*}
    &&(\phi^\ast f_2)(x\,;r,t) = \mathrm{Tr}_{V_x^{\vphantom{\perp}}}
    (r^2) + \mathrm{Tr}_{V_x^\perp} (t^2) \;, \\ &&(\phi^\ast f_1)
    (x\,;r,t) = 2\mathrm{i}\, \mathrm{Tr}_{V_x^{\vphantom{\perp}}}
    (A \Pi_x\, r) + 2\mathrm{i}\, \mathrm{Tr}_{V_x^\perp}
    (A (\mathrm{Id} - \Pi_x)\, t) \;.
\end{eqnarray*}
Next we fix $t \in D(V_x^\perp)$ and restrict the range of the
variable $r \in \mathrm{Sym}_B(V_x)$ to $E_x \cup F_x\,$. With this
restriction we get
\begin{eqnarray*}
    &&(\phi^\ast f_2)(x\,; \cdot,t) \big\vert_{E_x \cup F_x} =
    2\lambda^2 + \mathrm{const} \;,\\ &&(\phi^\ast f_1)(x\,; \cdot,t)
    \big\vert_{E_x} = 2\mathrm{i}\lambda \mathrm{Tr}_{V_x} (A)+
    2\mathrm{i} \eta\, B(e_x^+ , A e_x^+) + \mathrm{const}\;, \\
    &&(\phi^\ast f_1)(x\,; \cdot,t) \big\vert_{F_x} = 2\mathrm{i}
    \lambda \mathrm{Tr}_{V_x} (A) + 2\mathrm{i} \xi\, B(e_x^- ,
    A e_x^-) + \mathrm{const}\;.
\end{eqnarray*}
The coefficients $B(e_x^\pm , A e_x^\pm)$ never vanish. Indeed, from
$A = \sum_{a=1}^N \varphi_a \, B(\varphi_a \, , \cdot)$ we have
\begin{displaymath}
    B(e_x^\pm , A e_x^\pm) = \sum_{a=1}^N
    B(\varphi_a \, , e_x^\pm)^2 > 0\;,
\end{displaymath}
by the positivity assumption $A s > 0$ and the non-degeneracy of $B$.
As a consequence of the fact that $B(e_x^\pm , A e_x^\pm) \not= 0$,
it will suffice for our purpose of taking the limit $\varepsilon \to
0$ to examine the cutoff function $R\mapsto \chi_\varepsilon(R +
\mathrm{i} A)$ at $A = 0\,$.

By the relations $s e_x^+ = e_x^-$ and $s e_x^- = e_x^+$ the operator
$s$ commutes with the projector $\Pi_x = e_x^+ B(e_x^- , \cdot) +
e_x^- B(e_x^+ , \cdot)$. Therefore the pull back $\phi^\ast
\chi_\varepsilon$ of the cutoff function $R \mapsto \mathrm{e}^{-
\frac{\varepsilon}{2} \mathrm{Tr}\,(sR - Rs)^2}$ separates into two
factors, one for $V_x$ and $V_x^\perp$ each. We again keep $t \in
D(V_x^\perp)$ fixed and investigate the dependence on $r \in
\mathrm{Sym}_B(V_x)$. Using the easily verified relation $\Pi_x ( s r
- r s ) \Pi_x = (\xi(r) - \eta(r)) \Pi_x$ we obtain the expression
\begin{displaymath}
    (\phi^\ast \chi_\varepsilon)(x\,; \cdot , t) = \mathrm{e}^{ -
    \varepsilon (\xi - \eta)^2 + \mathrm{const}} \;.
\end{displaymath}

We finally conclude that $\int_\mathcal{E} \phi^\ast \omega$ vanishes
for $\omega = \mathrm{i}G_{A,\varepsilon} \, \iota_{\tau( \dot{A})}
|dR|$ in the limit $\varepsilon \to 0$. Indeed, we now see from the
above expressions that the inner integral over $E_x \cup F_x$ is
\begin{displaymath}
    \int_\mathbb{R} \mathrm{e}^{-2\lambda^2 - 2\mathrm{i}\lambda\,
    \mathrm{Tr}_{V_x}\, A} d\lambda \left( \int_\mathbb{R}\mathrm{e}^{
    -\varepsilon \eta^2 - 2 \mathrm{i}\eta B(e_x^+ ,\, A e_x^+)} d\eta
    + \int_\mathbb{R} \mathrm{e}^{-\varepsilon \xi^2 - 2 \mathrm{i}
    \xi B(e_x^- , \, A e_x^-)} d\xi \right) \;,
\end{displaymath}
and this always goes to zero when the regularization $\varepsilon$ is
sent to zero.

This shows that $\frac{d}{dt} I(A + t\dot{A}) \vert_{t = 0} = 0$ and
completes the proof of Thm.\ \ref{thm:1}.

\bigskip\noindent\textbf{Acknowledgment}. YVF acknowledges the financial
support of this research project by the Alexander von Humboldt
Foundation via a Bessel Research Award. The research in Nottingham
was supported by grant EP/C515056/1 from EPSRC (UK). The research of
MRZ is supported by the Deutsche Forschungsgemeinschaft, SFB/TR 12.

\end{document}